\documentclass[preprint,showpacs,preprintnumbers,amsmath,amssymb]{revtex4}

\usepackage{graphicx}
\usepackage{dcolumn}
\usepackage{bm}

\begin{document}
\title{Surface fraction of random close packing in two dimensions calculated exactly}

\author{   N.Olivi-Tran}
\affiliation{S.P.C.T.S., UMR-CNRS 6638, Universite de Limoges, 47 avenue Albert Thomas
87065 Limoges cedex, France}
\affiliation{G.E.S., UMR-CNRS 5650, Universite Montpellier II, 
case courrier 074, place Eugene Bataillon, 34095 Montpellier cedex 05, France
}
\date{\today}

\begin{abstract}
The random close packing fraction in two dimensions is calculated exactly via the
central limit theorem and the Brownian motion with no gain between monodisperse
disks.
\end{abstract}
\maketitle
\section{Introduction}
The value of the density of random close packing of granular matter is here exactly found 
for perfectly spherical monodisperse disks.
We use the central limit theory and the characteristics of Brownian
motion to demonstrate this. As a result we find, as commonly accepted
in the literature a value around 0.8 in 2 dimensions.
\section{Calculation}
Let us consider the Brownian motion of a particle and let us try to find it
at location $\bf R$ at time $t$, knowing that it is at the origin
at time $t=0$.
The central limit theorem allows one to obtain the distribution $R_N$
when $N$ tends towards infinity. This central limit theorem may be written
:
\begin{equation}
P({\bf R},t) \rightarrow \frac{1}{(4 \pi D t)^{d/2}}\exp -\frac{\bf R^2}{4Dt}
\end{equation}
with $t = N \tau$ and the diffusion constant $D=<{\bf r^2}>/(2 \tau)$

Let us now examine the scalar Brownian motion.
If $g$ is the gain at time $t$ knowing that it was zero at time $t=0$,
the probability to have a gain $g>0$ or a loss $g<0$ after a time $t$
(the time between two events is then taken as the unit interval) is \cite{fractale}:
\begin{equation}
P(g,t)=\frac{1}{\sqrt{2 \pi t}} \exp{-\frac{g^2}{2t}}
\end{equation}
The return to the origin (null gain) follows the law:
\begin{equation}
P(0,t)=\frac{1}{\sqrt {2 \pi t}}
\end{equation}

In two dimensions one can compare the Brownian motion with return to the origin
to the the number of particles visited by a random walker in a surface tending to infinity with a distance
between particles of the unit interval.
$t$ is then the sum of the distances $\ell$ between the centers of two touching particles (disks).
This distance $\ell$ is taken again as the unit distance. Therefore the surface fraction
will depend on  the number of visited particles.
 To obtain the number of particles
one has to calculate:
\begin{equation}
N_0=\int_O^t P(0,t')dt'=\int_0^t \frac{dt'}{\sqrt{2 \pi t'}}
=\sqrt{\frac{2}{\pi}}t^{1/2}
\end{equation}
The fact that we used a probability with gain equal to zero i.e. with return
to the origin ensures that all the particles have been
visited by the random walker.

This  gives a number of steps in  two dimensions equal to $\sqrt{\frac{2}{\pi}}t^{1/2}$.
In the last equation, overlaps between particles may occur 
thus we take the limit $t \rightarrow 1$ (the disk diameter
is assumed to be normalized and equal to 1).
This ensures that two particles do not overlap in the random walk, when
the random walker is visiting all particles. This is also equivalent
to a random walker walking with steps equal in length and of length equal to
1.
\begin{eqnarray}
N=\sqrt{\frac{2}{\pi}}t^{1/2}=\sqrt{\frac{2}{\pi}}\ell^{1/2}=
\\N(\ell=1)= \sqrt{\frac{2}{\pi}}=0.79 
\end{eqnarray}
with $\ell=1$ the length unit.

Equation(5) is the number $N$ of particles visited by the random walker
with unit steps.

Finally the surface fraction of $N$ monodisperse disks is then obtained
by equation (6) multiplied by 1, i.e. the normalized surface area of N disks.
The random close packing surface fraction is equal to approximatively 0.79 when the total
surface in which  the random close packing is embedded, is tending to $\infty$ to ensure
that all disks have been visited by the random walker. Furthermore, $N$ is a probability deduced from the central limit theorem, thus it is valid only in the
limit $r \rightarrow \infty$, where $r$ is related to the boundary conditions.

The fact that we  take the limit $\ell=1$  ensures that there are neither overlaps on the sides of the disks
nor on the normal direction regarding two disks.
In fact by taking  this limit we obtain a set
of non overlapping particles which have a maximum surface fraction.

Moreover the fact that we took a null gain ensures also non overlapping
of the disks indeed after having taken the limit $\ell \rightarrow 1$
the value of the maximum surface fraction does not depend any more on the length of the random walk.
But we have to add that our calculation is only valid for $r \rightarrow \infty$
as we used the central limit theorem which is only valid in this last case.
As a matter of fact the surface fraction depends on the boundary conditions.

\section{Conclusion}
We calculated exactly the value of the surface fraction of random close packing
in two dimensions.
In two dimensions, it is equal to $\sqrt{\frac{2}{\pi}}=0.79$ when
the number of monodisperse disks tends to $\infty$ but locally one may found
larger values of the random close packing density \cite{torquato}.
Indeed, computational works only deal with closed spaces which is not
the case here because we deal with an infinite plane.

This is obtained by a random walk with no gain (no back return) and with
random steps of length exactly equal to 1, i.e. with disk diameters normalized to 1.

\end{document}